\documentclass{article}

\usepackage[final]{nips_2017}

\usepackage[utf8]{inputenc} 
\usepackage[T1]{fontenc}    
\usepackage{hyperref}       
\usepackage{url}            
\usepackage{booktabs}       
\usepackage{amsfonts}       
\usepackage{nicefrac}       
\usepackage{microtype}      

\hypersetup{colorlinks=true,urlcolor=[rgb]{0.5, 0, 0},citecolor=[rgb]{0.5, 0, 0},linkcolor=[rgb]{0.5, 0, 0},filecolor=[rgb]{0.5, 0, 0}}

\title{Generalized Entropy Agglomeration}

\author{
  Dr. Işık Barış ~Fidaner\thanks{Computer Scientist, PhD. Admin to \href{https://yersizseyler.wordpress.com/kitap}{Yersiz Şeyler} [Placeless Things] website, Facebook groups \href{https://www.facebook.com/groups/Zatss/}{\textit{Žižek and the Slovenian School}} and \href{https://www.facebook.com/groups/subsets/}{\textit{Set Theory and Philosophy}}, and \href{https://twitter.com/MFSTurkey}{\textit{March for Science Turkey}} on Twitter. See: \href{https://fidaner.wordpress.com/science/}{fidaner.wordpress.com/science}. This is a technical report for REBUS 2.0.} \\
  Independent Computer Scientist\\
  \texttt{fidaner@gmail.com} \\
}

\usepackage{amsmath}
\DeclareMathOperator\gpei{gpei}

\begin{document}

\maketitle

\begin{abstract}
Entropy Agglomeration (EA) is a hierarchical clustering algorithm introduced in 2013. Here, we generalize it to define Generalized Entropy Agglomeration (GEA) that can work with multiset blocks and blocks with rational occurrence numbers. We also introduce a numerical categorization procedure to apply GEA to numerical datasets. The software \href{https://github.com/fidaner/entropy/tree/master/rebus2}{REBUS 2.0} is published with these capabilities: \begin{center}
\url{http://fidaner.wordpress.com/science/rebus2}
\end{center} 
\end{abstract}

\section{Introduction}

Entropy Agglomeration is a hierarchical clustering algorithm that was devised in [1] as part of an approach to a subproblem in Bayesian Nonparametrics. In using Chinese Restaurant Process for computing the posterior distributions of Dirichlet Processes, one encounters a fundamental question of combinatorial statistics: How can one summarize partitionings and feature allocations? Certain formulations were elaborated in [1] to represent the cumulative construction of partitioning structures, particularly the Cumulative Occurrence Distribution (COD),\footnote{Error correction for [1]: COD was misspelled as ``Cumulative Occurence Distribution'' in [1].} as well as the hierarchical clustering algorithm called Entropy Agglomeration (EA). The usage of EA was demonstrated on a famous literary text in [2] and the algorithm was implemented in Python under the name REBUS [3]. Here, we devise Generalized Entropy Agglomeration (GEA) by generalizing EA to be compatible with \textbf{feature allocations with multiset blocks} and \textbf{feature allocations with rational occurrence numbers}, as defined below. REBUS 2.0 is published with these capabilities [4].

\begin{enumerate}
\item In \textbf{partitionings and feature allocations} (as described in [1]), an element may occur only once in at least one block, or it may not occur in any block.\\e.g. $F=\{\{1,3,6,7\},\{2\},\{4,5\},\{5\}\}$ \\To illustrate, blocks may indicate the attendances of people to a meeting.\footnote{Intergovernmental organizations dataset in page 7-8 of [1] is a good example of this kind of interpretation.}
\item In \textbf{feature allocations with multiset blocks} (as described in Section 2), an element may occur once or multiple times in at least one block, or it may not occur in any block. \\e.g. $F=\{\{1,3,3,6,7\},\{2,2\},\{4,5\},\{5\}\}$ \\To illustrate, blocks may indicate amounts of participation by people in a meeting. 
\item In \textbf{feature allocations with rational occurrence numbers}  (as described in Section 3), the occurrence of an element in each individual block is indicated by a rational number up to a decimal digit, which becomes zero for the blocks where the element does not occur. Also a \textit{recurrence base} parameter $r$ is introduced and set as high as the larger occurrence numbers.\\e.g. $G=\{\{(1,
1.0),(3,2.0),(6,0.5)\},\{(2,2.1)\},\{(4,0.5),(5,0.3)\},\{(5,0.2)\}\}, r = 2.0$ \\To illustrate, blocks may indicate degrees of participation by people in a meeting.
\end{enumerate}

\section{Feature allocations with multiset blocks}

A multiset block of $[n]$ is a multiset that's composed by the elements of $[n]$. Here's an example:\footnote{Compare these examples to the illustrative example on the third page of [1].}
\begin{align*}
B\ =\ \{1, 3, 3, 6, 7\} 
\end{align*}

The size of a multiset block is the size of that multiset:
\begin{align*}
|B|\ =\ 1+2+1+1\ =\ 5
\end{align*}

A \textit{feature allocation with multiset blocks} of $[n]$ is a multiset of blocks $F=\{B_1,\dots,B_{|F|}\}$ where each $B_i$ is a non-empty multiset that's composed by the elements of $[n]$ for all $i\in\{1,\dots,|F|\}$.\footnote{Error correction for [1]: Two definitions in [1] on page 2 erroneously stated $i\in\{1,\dots,n\}$ whereas they should have been $i\in\{1,\dots,|Z|\}$ and $i\in\{1,\dots,|F|\}$.}

When multiset blocks are allowed in a feature allocation, the projection entropies computed from that feature allocation may turn out to have negative values. The three cases are:
\begin{enumerate}
\item When all elements occur exactly once in every block, entropy is zero.
\item When all elements occur at most once in every block, entropy is non-negative.
\item When elements may occur more than once in some blocks, entropy may be negative or non-negative.
\end{enumerate}

Non-negative entropies are convenient as they can be conceived as some kind of 'distance'. But when multiset blocks are allowed, entropies may become negative and it becomes harder to conceive the meaning of these computed quantities.

A feature allocation with multiset blocks is a special case of a feature allocation with rational occurrence numbers (described below) where the occurrence numbers can only be integers and the recurrence base is set to 1. It's sufficient for GEA to be compatible with rational occurrence numbers.

\section{Feature allocations with rational occurrence numbers}

A \textit{block with rational occurrence numbers} of $[n]$ is a multiset of 2-tuples, where each 2-tuple includes (1) an element in $[n]$ and (2) an occurrence number in $\mathbb{Q}$. Here is an example block with rational occurrence numbers:
\begin{align*}
B\ =\ \{(1, 1.0), (3, 2.0), (6, 0.5), (7, 0.3)\} 
\end{align*}

The size of a block with rational occurrence numbers is the sum of its occurrence numbers:
\begin{align*}
|B|\ =\ 1.0 + 2.0 + 0.5 + 0.3\ =\ 3.8
\end{align*}

A \textit{feature allocation with rational occurrence numbers} of $[n]$ is a multiset of blocks $G=\{B_1,\dots,B_{|G|}\}$ such that $B_i$ is a block with rational occurrence numbers of $[n]$ for all $i\in\{1,\dots,|G|\}$.

When blocks with rational occurence numbers are allowed in a feature allocation, the projection entropies computed from that feature allocation may turn out to have negative values (similar to the cases in Section 2). To reduce the negative entropies, a parameter called \textit{recurrence base} (defined in the next section) is introduced and set as high as the larger occurrence numbers in the feature allocation. The three cases are:
\begin{enumerate}
\item When all elements' rational occurrence numbers in every block is exactly equal to the recurrence base, entropy is zero.
\item When all elements' rational occurrence numbers in every block is not greater than the recurrence base, entropy is non-negative.
\item When some of the elements' rational occurrence numbers in some blocks are greater than the recurrence base, entropy may be negative or non-negative.
\end{enumerate}

\section{Generalized Entropy Agglomeration}

\textit{Generalized per-element information} is defined as (compare these to Equations 10 and 11 in [1]).
\begin{equation}
\gpei_n(B)\ =\ \int_{|B|}^{nr}\ \frac{1}{s}\ ds\ =\ \log\frac{nr}{|B|}
\end{equation}
where $r$ indicates the \textit{recurrence base}. \textit{Generalized Entropy} of a partitioning or feature allocation is:
\begin{equation}
H_g(G)\ =\ \sum_{i=1}^{|G|} \frac{|B_i|}{nr} \gpei_n(B_i) = \sum_{i=1}^{|G|} \frac{|B_i|}{nr} \log \frac{nr}{|B_i|}\ =\ \sum_{k=1}^{n}(\phi_k(G)-\phi_{k+1}(G))\frac{k}{nr}\log\frac{nr}{k}
\end{equation}

As a consequence of these definitions, Generalized Entropy Agglomeration (GEA) is simply an Entropy Agglomeration (EA) (as defined in [1]) that incorporates the \textit{recurrence base} in the equations.

In GEA, blocks that have larger rational occurence numbers may result in negative entropies, whereas a higher recurrence base may prevent the resulting negative entropies. However, a recurrence base that's too high causes GEA to output unbalanced dendrograms.

\section{Numerical categorization procedure to represent numerical data }

Since elements of $[n]$ function as categorically different indices, a feature allocation of $[n]$ can directly represent categorical data. However, to represent numerical data by a feature allocation, one needs a numerical categorization procedure. A simple numerical categorization procedure is introduced in this section:

Begin with an existing set of single dimensional numerical categories $\{x_j\}$ that may or may not belong to different dimensions.
\begin{itemize}
\item Take each $x_j$ as the central numerical category:
\begin{itemize}
\item Weight the central category with the value $w_j = 1$
\item Generate the neighborhood of the central category $x_j$:
\begin{itemize}
\item Generate the set of values $M = \{-m,-(m-1),\dots,-2,-1,1,2,\dots,m-1,m\}$ where $m$ is an integer overlap parameter.
\item For each $\mu\in M$:
\begin{itemize}
\item Add numerical category $x_{*} = x_j+\frac{\mu}{d}$ where $d$ is an integer division parameter.
\item Weight it with the value $w_{*} = (1-\frac{|\mu|}{m+1})^{\gamma}$ where $\gamma$ is a coefficient power.
\end{itemize}
\end{itemize}
\end{itemize}
\end{itemize}

After generating the neighborhoods for each central numerical category and weighting their elements, the numerical categories that have the exact same value on the same numerical dimension are connected by generating blocks with rational occurrence numbers:

For a set of numerical categories $\{x_1, x_2, \dots \}$ that have the exact same value on the same numerical dimension, generate a block with rational occurrence numbers $B = \{ (x_1, w_1), (x_2, w_2), \dots \}$.

An experimental numerical categorization was applied to the famous Iris dataset [5] by generating the neighborhood categories with parameters $d=10, m=5, \gamma=3$. GEA was applied to the resulting numerical categories with $r=1$. 145 of 150 flowers were correctly clustered by GEA.

A general description of the analysis procedure and a link to the software package is provided on the following webpage:

\begin{center}
\url{http://fidaner.wordpress.com/science/rebus2}
\end{center} 

\subsubsection*{Acknowledgments}

Thanks to Ali Taylan Cemgil from the Department of Computer Engineering in Boğaziçi University.

\section*{References}

\small

[1] I. B. Fidaner \& A. T. Cemgil. (2013) “Summary Statistics for Partitionings and Feature Allocations”. In Proceedings of \textit{Advances in Neural Information Processing Systems} (NIPS) 2013. Available on NIPS website.\\
\href{http://papers.nips.cc/paper/5093-summary-statistics-for-partitionings-and-feature-allocations}{papers.nips.cc}

[2] I. B. Fidaner \& A. T. Cemgil. (2014) “Clustering Words by Projection Entropy.” Poster accepted in Modern ML+NLP Workshop at NIPS 2014. \\
\href{https://fidaner.wordpress.com/science/rebus/}{fidaner.wordpress.com}, \href{https://arxiv.org/abs/1410.6830}{arxiv.org}, \href{http://www.cs.cmu.edu/~apparikh/nips2014ml-nlp/posters.html}{Workshop}

[3] Fidaner, I. B. \& Cemgil, A. T. (2014) REBUS 1.0: entropy agglomeration of text. Published under GNU General Public License. Online: \href{http://fidaner.wordpress.com/science/rebus/}{fidaner.wordpress.com/science/rebus}

[4] Fidaner, I. B. (2017) REBUS 2.0: entropy agglomeration of elements. Published under GNU General Public License. Online: \href{http://fidaner.wordpress.com/science/rebus2/}{fidaner.wordpress.com/science/rebus2}

[5] Fisher, R. A. (1936) The use of multiple measurements in taxonomic problems. \textit{Annals of Eugenics}, 7(2):179-188.\\ \relax

\end{document}